\documentstyle[epsfig,smacs,11pt]{article}
\title{ARTIFICIAL CATALYTIC REACTIONS IN 2D\\FOR COMBINATORIAL OPTIMIZATION}
\author{Jaderick P. Pabico}
\address{Institute of Computer Science, University of the Philippines Los Ba\~{n}os\\
   College 4031, Laguna, Philippines}
\email{jppabico@uplb.edu.ph}
\date{}

\abstract{Presented in this paper is a derivation of a 2D catalytic reaction-based model to solve combinatorial optimization problems (COPs). The simulated catalytic reactions, a computational metaphor, occurs in an artificial chemical reactor that finds near-optimal solutions to COPs. The artificial environment is governed by catalytic reactions that can alter the structure of artificial molecular elements. Altering the molecular structure means finding new solutions to the COP. The molecular mass of the elements was considered as a measure of goodness of fit of the solutions. Several data structures and matrices were used to record the directions and locations of the molecules. These provided the model the 2D topology. The Traveling Salesperson Problem (TSP) was used as a working example. The performance of the model in finding a solution for the TSP was compared to the performance of a topology-less model. Experimental results show that the 2D model performs better than the topology-less one.}

\begin{document}
\maketitle

\section{Introduction}

Solutions to combinatorial optimization problems (COPs) have practical real-world importance because most real-world problems are combinatorial in nature. Most COPs have been shown to be ${\cal NP}$-complete. Exact algorithms have been proposed to these problems but prove inefficient for large problem instances~\cite{garey79}. Graph-based algorithms such as branch and bound~\cite{tschoke95}, as well as distributed multi-agent based heuristics such as genetic algorithms~\cite{pabico99}, memetic algorithms~\cite{moscato92,freisleben96a,freisleben96b}, tabu search~\cite{zachariasen95}, simulated annealing~\cite{martin96}, simulated jumping~\cite{amin99}, neural networks~\cite{miglino94}, and swarm intelligence~\cite{gambardella95,gambardella96,dorigo97} have been used to find time-restrained optimal and near optimal solutions for these problems.

In recent years, the chemical systems of living organisms have been shown to possess inherent computational properties~\cite{hjemfelt91,adleman94,arkin94}. This discovery provided researchers the chemical metaphor as a paradigm for computation~\cite{berry92,fontana92,banzhaf95,ikegami95,banzhaf96,dittrich98}. Under this computational framework, molecules are considered as solutions, while interactions among molecules represent computational procedures.

Presented in this paper are the mapping of permutations to molecules, and the derivation of two stochastic functions that model catalytic reactions. The two functions simulate a unary catalytic reaction and a binary catalytic reaction. These reactions create new molecules where the average goodness of fit of the product is better than the average goodness of fit of the reactants. A molecule encodes a permutation such that the unary catalytic reaction reorders the element in the permutation while the binary catalytic reaction creates new permutations. 

\section{Model Development}

This section discusses the development of the 2D artificial catalytic reactor (2DACR). The 2DACR models the dynamics of artificial molecules in a 2D environment. The environment is driven by several rules of interactions to produce a set of optimal or near-optimal solutions to a COP. The 2DACR is defined by a triple ${\rm 2DACR}(M, R, A)$, where $M$ is a set of artificial molecules, $R$ is a set of reaction rules describing the interaction among molecules, and $A$ is an algorithm driving the reactor. In this paper, the molecules in $M$ are permutations while the rules in $R$ are reordering algorithms that create new molecules. The algorithm $A$ describes how the rules are applied to a vessel of artificial molecules simulating a well-stirred, 2D reactor. Further, the 2DACR is partitioned by $A$ into different levels of reaction activities. The level of reaction activity is a function of molecular mass. 

\subsection{Mapping 2DACR to COP}

In a particular COP, if $n$ is the length of a permutation, then all $n$-sized permutations $\Pi_n$ are molecules. The inherent value of a permutation $\pi\in\Pi_n$ is the mass of the molecule and is computed by the objective function of the COP. For example, if the COP is a traveling salesperson problem (TSP) with a set of cities $V$, a set of paths $E$,and a cost matrix $G$, then all permutations of the $|V|$ cities are molecules. Each $n$-sized molecule is an encoding of a Hamiltonian tour such that the ordering of the cities $v_i\in V$ represents a molecular structure. Each city $v_i$ is a distinct atom in the molecule. The cost $f_v$ of traversing a Hamiltonian tour is the mass of the molecule and is a measure of goodness of fit of the Hamiltonian tour. If the objective function is a minimization of the tour cost, then light molecules encode the desirable solutions. In this example, $f_v$ is defined in Eq.~(\ref{eqn:cost}) where $g_{i,j}\in G$ is the cost measure associated with path $(v_i,v_j)\in E$:

\begin{equation}
f_v = g_{n,1} + \sum_{i=1}^{n-1} g_{i, i+1}.\label{eqn:cost}
\end{equation}

\subsection{Artificial Catalytic Reactions}

If two molecules $m_1$ and $m_2$ collide, they react following a catalytic reaction of the form \[m_1 + m_2 + {\cal C} \longrightarrow m_1 + m_2+ m_3 + m_4,\] where $m_3$ and $m_4$ are product molecules and ${\cal C}$ is a catalyst. The reaction is a mathematical function \[R_1: M\times M \longrightarrow M\times M\times M\times M,\] where $M = \{m_i\quad\forall i=1,2,\dots,|\Pi_n|\}$, $m_i$ is a molecule and is a linear data structure representing a permutation, and $|\Pi_n|$ is the cardinality of the solution space of the COP (i.e., $|M|=|\Pi_n|$). $R_1$ performs a reordering of solutions and is dependent on an $n$-long vector $S$ with binary elements. The elements $s\in S$ are computed following this algorithm:

\begin{enumerate}
\item Let ${\rm random}(n)$ be a function that returns a random integer between 1 and $n$.
\item Let ${\rm index}(m_1, m_{2,j})$ be a function that returns the position of atom $m_{2,j}$ in molecule $m_1$.
\item Let $\omega_1$ be a set of distinct atoms in molecule $m_1$.
\item Let $m_{1,j}$ be the $j$th atom in molecule $m_1$.
\item Let the integers $i$ and $j$ be indeces
\item Initialize the vector $S=0$.
\item Initialize $\omega_1=\{\}$.
\item Set $i=0$.
\item Let $j={\rm random}(n)$ be the point of collision between $m_1$ and $m_2$.
\item Do the following until $i=n$ or $m_{2,j}\in\omega_1$:
  \begin{enumerate}
  \item Increment $i$ by one.
  \item Set $\omega_1 = \omega_1 \bigcup \{m_{1,j}\}$.
  \item Set $s_j = 1$.
  \item If $m_{2,j} \not\in \omega_1$, then $j={\rm index}(m_1, m_{2,j})$.
  \end{enumerate}
\end{enumerate}

Once $S$ is obtained, the molecules $m_3$ and $m_4$ can be computed using the following algorithm:

\begin{enumerate}
\item For each $i=1, \dots, n$ do:
  \begin{enumerate}
  \item If $s_i = 1$, then set $m_{3,i}=m_{1,i}$ and $m_{4,i}=m_{2,i}$;
  \item Else set set $m_{3,i}=m_{2,i}$ and $m_{4,i}=m_{1,i}$.
  \end{enumerate}
\end{enumerate}

If a molecule $m_5$ hits the bottom or walls of the reactor, a catalytic reaction of the form \[m_5+{\cal C}\longrightarrow m_6\] happens. The reaction is a mathematical function \[R_2:M\longrightarrow M\] described by the following algorithm:

\begin{enumerate} 
\item Let $j={\rm random}(n)$ be the point of collision between $m_5$ and the reactor wall or bottom.
\item Let $i={\rm random}(2)-1$.
\item If $i=0$, then do the following:
  \begin{enumerate}
  \item If $j=n$, then $m_{6,1}=m_{5,j}$,
  \item Else $m_{6,j+1}=m_{5,j}$.
  \end{enumerate}
\item If $i=1$, then do the following:
  \begin{enumerate}
  \item If $j=1$, then $m_{6,n}=m_{5,j}$,
  \item Else $m_{6,j-1}=m_{5,j}$.
  \end{enumerate}
\end{enumerate}

\subsection{Two-Dimensional Reactor}

The reactor algorithm $A$ operates on a matrix $T$ of molecules and catalysts, where $T=\{t_{i,j} | t_{i,j}=m_k\vee t_{i,j}=0\quad\forall i=1,\dots,I \wedge j=1,\dots,J\}$, $I>1$, $J>1$ and $(I\times J)<<|M|$. $m_k$ is the $k$th molecule while the catalyst is when $t_{i,j}=0$. Here, the matrix element $t_{i,j}$ serves a placeholder for the molecule $m_k$ such that the expression $t_{i,j}=m_k$ should be understood as ``$m_k$ is at $t_{i,j}$.'' 

\subsection{Molecule Direction}

\begin{figure}
\centerline{\epsfig{file=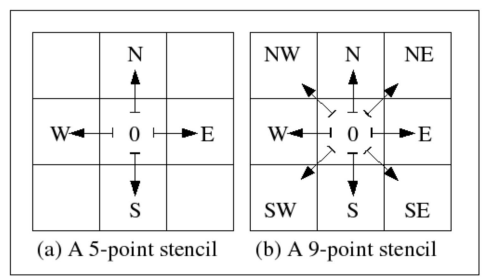,width=2.5in,height=1.42in}}
\caption{Two of the possible direction stencils for a 2D environment.}
\label{fig:stencil}
\end{figure}

Associated with each molecule $m_k$ is a direction $d_k$ of the molecule. The range of values for $d_k$ is dependent on the {\em stencil} used by $A$. A {\em stencil} is a set of directions from an element in a grid environment. In a 2D environment, the possible stencils are 5-point stencil and 9-point stencil (Figure~\ref{fig:stencil}). There are five possible directions in a 5-point stencil: no movement (0), due North (N), due East (E), due West (W), and due South (S). In a 9-point stencil, the nine possible directions are 0, N, NE, E, SE, S, SW, W, and NW. An arbitrary integer may be assigned to each of these directions reserving the value zero for the {\em no movement} direction. Elements at the borders and at the corners also follow any of these stencils.

A random direction is assigned to each molecule during the start of the simulation. After a collision of type $R_1$ and following a 9-point stencil, the directions of the products are determined as follows:
\begin{enumerate}
\item If the mass of a molecule is greater than the average mass of the products, then the direction assigned is ${\rm random(NW, N, NE)}$.
\item If the mass of a molecule is lesser than the average mass of the products, then the direction assigned is ${\rm random(SW, S, SE)}$.
\item If the mass of a molecule is the same\footnote{In practical application, this may mean as {\em statistically the same}.} as the average mass of the products, then  the direction assigned is ${\rm random(W, 0, E)}$.
\end{enumerate}

After a collision of type $R_2$ and following a 9-point stencil, the direction of the product is determined as follows:
\begin{enumerate}
\item If the mass of the product is greater than the mass of the reactant, then the direction assigned is ${\rm random(NW, N, NE)}$.
\item If the mass of the product is greater than the mass of the reactant, then the direction assigned is ${\rm random(SW, S, SE)}$.
\item If the mass of the product is the same as the mass of the reactant, then the direction assigned is ${\rm random(W, 0, E)}$.
\end{enumerate}

\subsection{Collision Matrix}

The algorithm $A$ also maintains a collision matrix $C$ whose elements are defined by Eq.~(\ref{eqn:collision}). The matrix $C$ has the same dimension as $T$ and records which molecules will collide at which element in $T$. Two molecules $m_k$ and $m_l$ will collide at $t_{i,j}$ if $c_{i,j}=(m_j, m_k)$. The collision will obey the reaction rule defined by $R_1$. A molecule $m_k$ will collide with the border at $t_{i,j}$ if $c_{i,j}=m_k$. In this case, the reaction rule $R_2$ will be applied. If $c_{i,j}=0$, then no collision will occur at $t_{i,j}$. Each element $c_{i,j}\in C$ is updated by tracing the movement of a molecule via its direction. Usually, $C$ is updated via any of the two popular computational order: row-major ordering and column-major ordering. However, these two popular ordering techniques have inherent biases. The row-major ordering scheme is biased towards molecules at lower values of $i$ (i.e., at the upper portion of $T$) while the column-major ordering is biased towards those at lower values of $j$ (i.e., at the left portion of $T$). To remove the biases brought about by these two ordering schemes, ordering schemes based on space-filling curves are recommended. In this paper, the Morton ordering scheme is used as a working example and is discussed in the next section.

\begin{equation}
  c_{i,j}=\left\{\begin{array}{cl}
          (m_k, m_l) & {\rm if\ } m_k \wedge m_l {\rm\ collide\ at\ } t_{i, j}\\
	  m_k & {\rm if\ } m_k {\rm\ collides\ with\ the\ wall\ at\ } t_{i,j}\\
	  0 & {\rm if\ no\ collision\ occurs\ at\ } t_{i,j}
          \end{array}\right.\label{eqn:collision}
\end{equation}

\subsection{Ordering of Computation}

\begin{figure}
\centerline{\epsfig{file=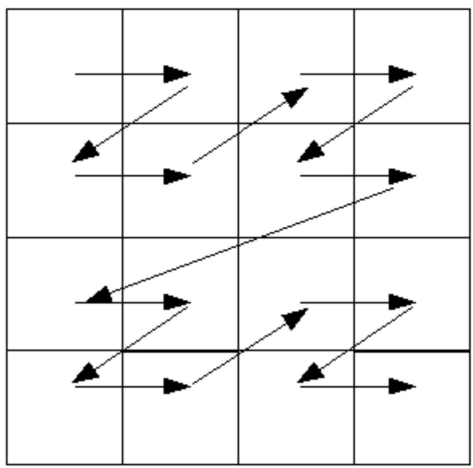,width=2.5in,height=2.44in}}
\caption{Order of computation for a $4\times 4\quad T$ following the Morton ordering scheme.}
\label{fig:ordering}
\end{figure}

The Morton ordering (Figure~\ref{fig:ordering}) is an ordering scheme based on Peano-Hilbert plane-filling curves. A plane-filling curve is a curve drawn on a plane and fills it. A plane-filling curve is efficiently generated using a recursive algorithm that divides the plane at each recursion level. The principle followed at each recursion is the self-similarity principle such that the structure of the curve at the higher level is the same as the structure of the curve at the lower level. Traversing a curve means to enumerate the points along the curve. When the curve fills the plane, the traversal of the curve also means the traversal of the plane at the order defined by the curve. The order of traversal is defined by a {\em universal turtle traversal algorithm}~\cite{jin05}, which traverses a self-similar space-filling curve based on a movement specification table.

\subsection{Evolution of the Reactor}

The evolution in $A$ is realized by applying the following algorithm:
\begin{enumerate}
\item Initialize $T$ with $I\times J$ molecules selected randomly from $M$.
\item Compute the molecular mass of each $m_k\in T$ and initialize each with a direction.
\item Compute for $C$ using the Morton ordering scheme.
\item Apply the reaction rule $R_1$ for any $m_1$ and $m_2$.
\item Apply the reaction rule $R_2$ for heavy molecules that collide with the reactor walls and bottom. 
\item Decay the heavier molecules by removing them out of $T$ and replacing them with randomly selected molecules from $M$.
\item Repeat steps~2 to~5 until $T$ is saturated with lighter molecules.
\end{enumerate}

One iteration of $A$ constitutes one epoch in the artificial reactor. The  sampling procedure gives molecules with low molecular mass a higher probability to react or collide with other molecules. This mimics the level of excitation energy the molecule needs to overcome for it to react with another molecule. This means that the lighter the molecule, the higher the chance that it will collide with other molecules. Step~6 of algorithm $A$ requires a metric for measuring saturation of molecules. The reactor is considered saturated when $T$ has no more 0 element (i.e., the catalyst is already exhausted).

\section{Experimental Results}

A 2DACR was run to solve an instance of a symmetric TSP. The 2DACR used the 5-point stencil and the Morton ordering scheme. To assess the performance of the 2DACR, a topology-less artificial chemical reactor (0DACR) recently employed by other researchers~\cite{pabico04,pabico06} was also run to solve the same TSP instance. A single-processor Pentium IV machine with 1.2GHz bus speed running under a multiprogramming operating system was used to run the 2DACR and the 0DACR simulations. The simulations were repeated 10 times while the best minimum Hamiltonian tour length for each run were recorded. The values recorded were averaged and the standard deviation computed. Shorter tour lengths imply better tour costs and are much desirable. To remove the initialization bias, both 2DACR and 0DACR used the same initial set of 500 molecules, with the exception that the molecules in 2DACR have their respective initial locations and directions while those of the 0DACR have none. The 2DACR utilized a $30\times 30$ matrix~$T$ with 400 elements in~$T$ acting as the catalyst~${\cal C}$. Table~\ref{tab:stsp} compares the average tour lengths found by 2DACR and 0DACR on five sets of random instances of symmetric 50--city TSPs. The table shows the average value of 10 runs for both 2DACR and 0DACR with their respective standard deviations. Based on the result presented, it can be seen that 2DACR's performance is better than the performance of the 0DACR employed by other researchers.

\begin{table}[htb]
\caption{Comparison of average Hamiltonian tour length found by 2DACR and 0DACR on five sets of random instances of symmetric 50--city TSPs. The values are averaged over 10 runs. The values in parenthesis are the standard deviations.} \label{tab:stsp}
\centering
\begin{tabular}{cccc}
\hline\hline
Problem & 0DACR (std. dev.) & 2DACR \\
\hline
1&	5.89 (0.40)&	5.87 (0.45)\\
2&	6.17 (0.08)&	6.15 (0.11)\\
3&	5.65 (0.21)&	5.59 (0.18)\\
4&	5.70 (0.98)&	5.67 (0.77)\\
5&	6.15 (0.54)&	6.14 (0.55)\\
\hline\hline
\end{tabular}
\end{table}

\section{Concluding Remarks}

An algorithm that models catalytic reactions in 2D was designed to solve COPs using the TSP as a working example. Solutions to an instance of TSP via 2DACR were found better on the average than those found by 0DACR. Molecular directions and locations were incorporated that provide the model with a 2D topology. The order of computation via Morton ordering might have removed the bias inherent in row-major and column-major ordering schemes. However, more work is needed that will compare the performance of 2DACR using these ordering schemes.

\section*{Acknowledgments}

This work is funded by the Institute of Computer Science, College of Arts and Sciences, University of the Philippines Los Ba\~{n}os. 

\bibliographystyle{abbrv}
\bibliography{smacs06}

\end{document}